\documentclass[journal,comsoc]{IEEEtran}
\usepackage[T1]{fontenc}% optional T1 font encoding
\usepackage{textcomp}
\usepackage{graphicx}

\ifCLASSINFOpdf
\else
\fi

\usepackage[cmex10]{amsmath}
\usepackage{amssymb}   % \triangleq 
\usepackage{amsxtra}
\usepackage{amscd}
\usepackage{amsthm}
\usepackage{textcomp}
\usepackage{graphicx}  
\usepackage{balance}
\usepackage{array}  
\usepackage{tablefootnote}
\usepackage{multirow}  
\usepackage{amsmath}
\usepackage{cite}
\usepackage{times}
\usepackage{lipsum}
\usepackage{footnote}
\usepackage{booktabs, makecell}
\usepackage{fixltx2e}
\usepackage[flushleft]{threeparttable}
\usepackage{blindtext}
\usepackage[cmintegrals]{newtxmath}
\usepackage{url}
\usepackage{color,soul} % for highlights
\soulregister\cite7
\soulregister\ref7
\soulregister\pageref7
 
\usepackage{multirow}

\begin{document}

\title{Reconfigurable Intelligent Surfaces for Future Wireless Networks: A Channel Modeling Perspective}

\author{Ertugrul~Basar,~\IEEEmembership{Senior Member,~IEEE} and
        Ibrahim~Yildirim,~\IEEEmembership{Graduate Student Member, IEEE}
        % <-this % stops a space
	\thanks{E. Basar and I. Yildirim are with the Communications Research and Innovation Laboratory (CoreLab), Department of Electrical and Electronics Engineering, Ko\c{c} University, Sariyer 34450, Istanbul, Turkey. e-mail: ebasar@ku.edu.tr}
		\thanks{I. Yildirim is also with the Faculty of Electrical and Electronics Engineering, Istanbul Technical University, Istanbul 34469, Turkey. e-mail: yildirimib@itu.edu.tr}
	
			\thanks{Codes available at https://corelab.ku.edu.tr/tools/SimRIS }}

\markboth{ }%
{Shell \MakeLowercase{\textit{et al.}}: SimRIS MIMO}

\maketitle

\begin{abstract}
While the researchers have set their sights on future wireless networks of 2030, communications through reconfigurable intelligent surfaces (RISs) appears as one of the potential enabling technologies for 6G wireless networking. This article aims to shed light on the potential use-cases of RISs in future wireless systems by means of a novel channel modeling methodology as well as a new software tool for RIS-empowered millimeter-wave communication systems. It is shown by the open-source, user-friendly, and widely applicable \textit{SimRIS Channel Simulator}, whose 2.0 version is proposed and goes online by this article, that RISs will work under certain use-cases and communication environments. Potential future research directions are also discussed to bridge the gap between the theory and practice of RIS-empowered systems towards their standardization for 6G wireless networks. 
\end{abstract}

\begin{IEEEkeywords}
6G, channel modeling, MIMO systems, reconfigurable intelligent surfaces, SimRIS Channel Simulator. 
\end{IEEEkeywords}

\IEEEpeerreviewmaketitle

\section{Introduction}

\begin{figure*}[!t]
	\begin{center}
		\includegraphics[width=1.9\columnwidth]{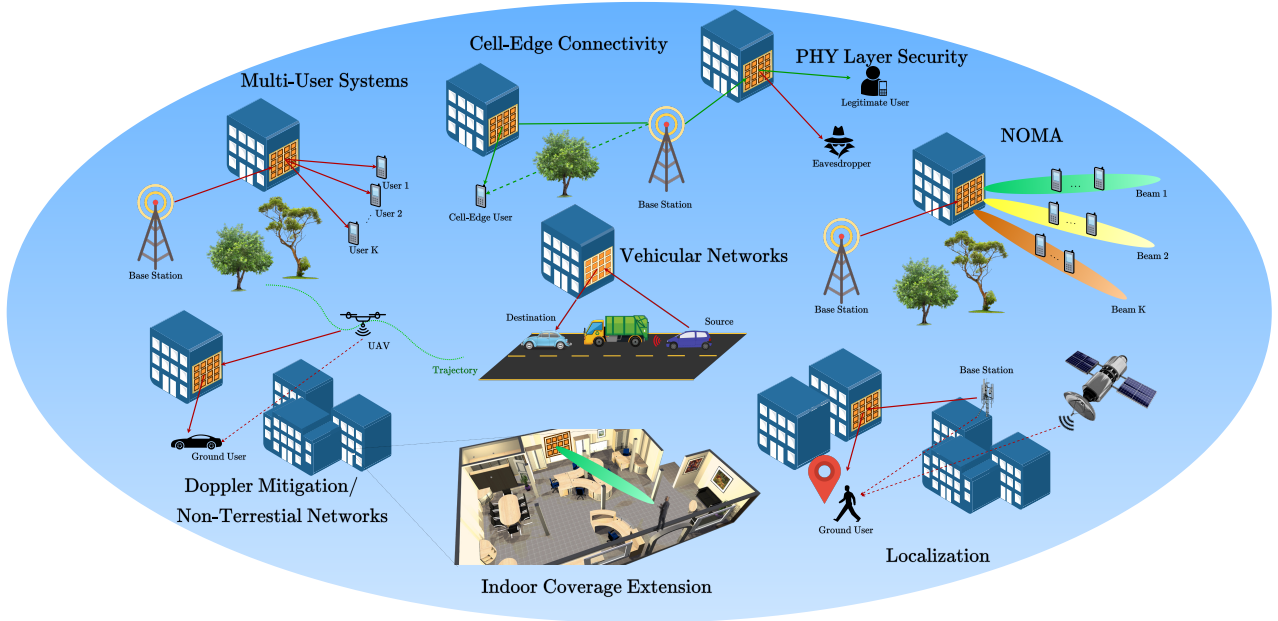}
		\vspace*{-0.1cm}\caption{Emerging applications and use-cases of RIS-assisted communication systems in future wireless networks.}\vspace*{-0.5cm}
		\label{fig:App}
	\end{center}
\end{figure*}

\IEEEPARstart{T}{he} wireless community is witnessing very exciting period in which the standardization and commercialization of 5G has been completed, and academia and industry have set their sights on 6G wireless technologies of 2030 and beyond by starting active research campaigns. According to July 2020 white paper of Samsung Research  \cite{Samsung_2020}, three key services will dominate 6G: truly immersive extended reality (XR), high-fidelity mobile hologram, and digital replica. These interesting and advanced multimedia services will require a peak data rate of 1,000 Gbps and a user experienced data rate of 1 Gbps with an air latency less than 0.1 ms. These highly challenging performance  requirements of 6G urged the researchers to investigate alternative solutions in the physical layer (PHY) of future wireless networks. Within this context, promising solutions such as energy harvesting, optical wireless communications, communications with intelligent surfaces, aerial communications, cell-free massive multiple-input multiple-output (MIMO) systems, full-duplex communications, and Terahertz communications have been put forward in recent years \cite{Rajatheva_6G}. At this point, the concept of communications with \textit{reconfigurable intelligent surfaces (RISs)} stands out as a strong candidate due to its undeniable potential in realizing truly pervasive networks by re-engineering and recycling the electromagnetic (EM) waves.

It is an undeniable fact that a level of saturation has been reached in current wireless systems despite the use of sophisticated PHY techniques such as adaptive modulation and coding, multi-carrier signaling, large-scale MIMO systems, relaying, beamforming, reconfigurable antennas and so on. As a result, next-generation wireless systems require a radical transformation to satisfy the unprecedented demands of 6G networking by enabling software control, not only in the network layer as in 5G but also at various layers of the communications stack. Within this context, RIS-assisted communication challenges the current status quo in legacy wireless systems by providing means to manipulate the wireless signals over the air \cite{Akyildiz_2018,Basar_Access_2019}. This is accomplished by RISs, which are man-made and two-dimensional meta (\textit{beyond} in ancient Greek)-surfaces equipped with integrated electronics to enable reconfigurable EM properties. 

RIS-empowered communication systems have received tremendous interest from academia and industry in recent times and have been explored in numerous applications, including but limited to multi-user systems, PHY security, low-cost transmitter implementation, vehicular and non-terrestrial networks, non-orthogonal multiple access, device-to-device communication, Internet-of-things (IoT), Doppler and multi-path mitigation, localization and sensing, and so on \cite{Wu_Tutorial}. Despite their rich use-cases shown in Fig. 1, in which the intelligent scattering feature of RISs is exploited to improve the end-to-end system performance, potential killer applications of RISs have not been well understood in the open literature. This can be mainly attributed to the lack of practical and physical RIS-assisted channel models for different operating frequencies and bands, which considers physical characteristics of RISs.

Against this background, this article takes a step back and approaches RIS-empowered communication systems from the perspective of physical channel modeling to shed light into their practical use-cases in future networks. Building on our previous channel modeling campaign presented in \cite{SimRIS_jour_new} and \cite{SimRIS_Latincom}, in this article, we also propose \textit{SimRIS Channel Simulator v2.0} by incorporating MIMO-aided transmit and receive terminals into its earlier version. Our comprehensive channel modeling efforts that include 3D channel models for 5G networks and physical RIS characteristics, and computer simulations reveal that \textit{RISs will work!} by carefully adjusting their size, position, and functionality. In other words, using a physical channel modeling methodology that is unique to RIS-empowered systems, we prove that the received signal quality, in return the achievable rate, can be significantly boosted for both indoor and outdoor environments at certain locations. 

It will still take a village to find energy- and cost-efficient ways to reconfigure practical RISs in real time with fully passive elements, however, this article may help to validate the potential applications of emerging RISs in 6G and beyond wireless networks. We hope that this article, which follows a novel channel modeling perspective, might be a useful source for future practical implementation campaigns with sophisticated RISs and can be supported with real data in the future. The open-source nature of the SimRIS Channel Simulator might also be very helpful for interested researchers and engineers, and the outcomes (generated channels) of our simulator can be easily used for the evaluation (capacity, secrecy, error/outage performance etc.)  of existing RIS-assisted systems. This article also reviews the state-of-the-art use-cases and potential future directions of RIS-empowered networks and might be a useful source for interested readers.

\section{Reconfigurable Intelligent Surfaces: How Do They Work?}

RISs are man-made surfaces of EM material that are electronically controlled with integrated electronics, and have unnatural and unique wireless communication capabilities. The core functionality of RISs relies on a basic physics principle, which states that EM emissions from a surface are defined by the distribution of electrical currents over it. The surface currents are caused by the impinging EM waves and RISs aim to control and modify the current distribution over them in a deliberate manner to enable exotic EM functionalities, including wave absorption, anomalous reflection, and reflection phase modification. This control is accomplished by ultra-fast switching elements such as varactors, PIN diodes, or micro-electro-mechanical systems (MEMS) switches to manipulate the available RIS elements. In different terms, software-defined RISs that are placed in propagation environments, enable wireless network designers to control the scattering, reflection, and refraction characteristics of radio waves, by overcoming the negative effects of natural wireless propagation.

A typical RIS consists of three main layers. The outermost layer consists of a large number of small and metallic (copper etc.) patch elements printed on a dielectric material and interacts directly with the incoming signals. The second layer is a conducting back-plane that prevents the energy leakage to the back of the RIS. Finally, the third layer contains control circuits connected to a micro-controller, which distinguishes an RIS from a passive reflect-array. The micro-controller might be equipped with an IoT gateway to receive commands from a central controller or the base station to reconfigure the RIS. Alternatively, it might run artificial intelligence (AI) algorithms in a stand-alone mode to determine the states of the RIS. Although the operation of the RIS itself is passive, that is, without requiring a power source for signal reflection, the RIS needs a power source for its micro-controller and switch elements.

To illustrate the concept of communications over an RIS with $N$ tiny (sub-wavelength) reflecting elements, let us consider the basic use-case of cell-edge connectivity in Fig. 1, where the direct path between the transmitter (Tx) and the receiver (Rx) is blocked by obstacles. Denoting the distances between the Tx and the RIS and the RIS and the Rx by $d_1$ and $d_2$ respectively, according to the free-space propagation and plate scattering theory, the received (maximized) signal power $P_r$ follows 
\begin{equation}
  P_r \propto P_t \frac{N^2\lambda^4}{(4\pi)^2 d_1^2 d_2^2}.  
\end{equation}
Here, $P_t$ is the transmit power, $\lambda$ is the wavelength, and the gains of antennas and RIS elements are ignored for brevity. We further assume that the variation of $d_1$ and $d_2$ is negligible across the RIS, which is a valid assumption in the far-field. We further note that $P_r$ is always less than $P_t$ due to the conservation of power and we have different power scaling laws for the near-field as recently reported with practical measurements in \cite{Tang_2020}.

Considering the above equation, our observations are as follows: First, scattering is not merciful as received signal power decays with $d_1^2 d_2^2$, which requires  very careful planning and positioning for the RIS. In other words, the Rx might not even notice the existence of an RIS that is relatively far away from either the Tx or the Rx. Accordingly, the received signal can only be  boosted significantly when the LOS path is not strong enough or fully blocked. Second, thanks to phase coherent combining of all scattered signals from the RIS, which is possible by careful adjustment of phase responses (time delays or impedances) of RIS elements, we have the term of $N^2$ in the received signal power. This requires a careful planning on the overall size, total cost, and physical structure of the RIS. Although increasing $N$ is a great booster to the overall system performance, it also increases the cost, complexity, and training overhead of RISs in return. At this point, we might use AI-aided tools or sophisticated optimization methods to reach a compromise.

Finally, it is worth noting that communication through intelligent surfaces is different compared with other related technologies currently employed in wireless networks, such as relaying, MIMO beamforming, passive reflect-arrays, and backscatter communications. This can be attributed to the unique features of RISs such as their nearly passive structure without requiring any RF signal processing, up/down-conversion, power amplification or filtering, inherently full-duplex nature, low cost, and more importantly, real-time reconfigurability. 

These distinctive characteristics make RIS-assisted communication a unique technology and introduce important design challenges that need to be tackled under the rich applications and use-cases in Fig. 1. Our channel modeling campaign and SimRIS Channel Simulator consider physical millimeter-wave (mmWave) channels and RIS characteristics to provide useful insights to system designers.

\begin{figure*}[!t]
	\begin{center}
		\includegraphics[width=1.9\columnwidth]{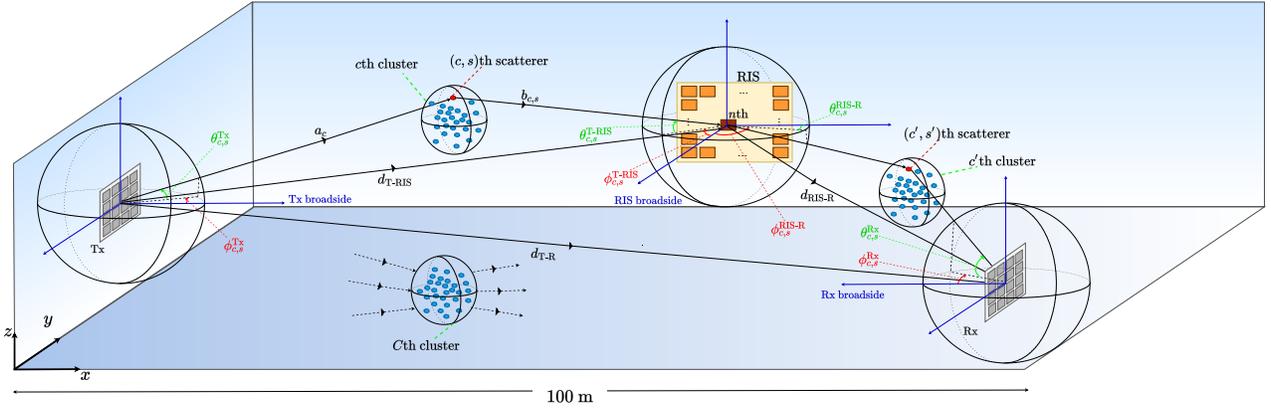}
		\vspace*{-0.3cm}\caption{Generic 3D communication environment with random clusters/scatterers and an RIS mounted on the $xz$-plane.}\vspace*{-0.1cm}
		\label{fig:Fig1}
	\end{center}
\end{figure*}

\section{Channel Modeling for RIS-Empowered Systems}
Despite their rich applications and potential use-cases in future wireless networks, there is still an urgent need for physical, open-source, and widely applicable channel models involving RISs to be used in various
systems in indoor and outdoor environments at different operating frequencies. Considering that RIS channel modeling is the first step towards future RIS-empowered networks, the importance of accurate and physical RIS-assisted channel models even increases. This article aims to shed light on this fundamental issue by laying foundations for physical channel modeling of RIS-empowered communications.

The scattering nature of RISs in the far-field, their reconfigurable amplitude and phase responses, and fixed orientation render channel modeling for RIS-based systems an interesting problem. Due to the scattering nature of RISs, each RIS element behaves as a new signal source by capturing the transmitted signals \cite{SimRIS_jour_new}. Let us consider a MIMO system with $N_t$ transmit and $N_r$ receive antennas operating in the presence of an RIS with $N$ reflecting elements. Consequently, the end-to-end channel matrix $\mathbf{C} \in\mathbb{C}^{N_r \times N_t}$ of an RIS-assisted MIMO system can be given as follows
\begin{equation}
\mathbf{C}=\mathbf{G} \mathbf{\Phi} \mathbf{H} + \mathbf{D}. 
\end{equation}
Here, $\mathbf{H}\in\mathbb{C}^{N \times N_t}$ is the matrix of channel coefficients between the Tx and the RIS, $\mathbf{G}\in\mathbb{C}^{N_r \times N}$ is the matrix of channel coefficients between the RIS and the Rx, and $\mathbf{D} \in\mathbb{C}^{N_r \times N_t}$ stands for the direct channel (not necessarily a line-of-sight (LOS)-dominated one and is likely to be blocked for mmWave bands due to obstacles in the environment) between the Tx and the Rx. The beauty of the RIS-assisted model lies in the middle of the composite channel: $\mathbf{\Phi}$, which is a diagonal matrix capturing the (complex) reconfigurable responses of the RIS elements. The researchers have been working on the optimization of $\mathbf{\Phi}$ since the early works of \cite{Wu_2018_3} and \cite{Huang_2019}  under different constraints and objectives: maximizing the signal-to-noise ratio (SNR), capacity, sum-rate, or secrecy and reducing the interference. In a physical channel model for RIS-assisted systems, the LOS probability between terminals as well as the gain and array response of RIS elements should be carefully reflected to channel matrices of $\mathbf{H}$ and  $\mathbf{G}$. Furthermore, the close proximity of the RIS to the terminals, its limited field of view and 2D architecture as well as fixed orientation complicates RIS channel modeling compared to simplified cascaded channel models and brings new technical challenges. As we will discuss next, SimRIS Channel Simulator aims to open a new line of research by taking into account all these effects in channel modeling.

\section{SimRIS Channel Simulator}

\begin{figure}[!t]
	\begin{center}
		\includegraphics[width=0.95\columnwidth]{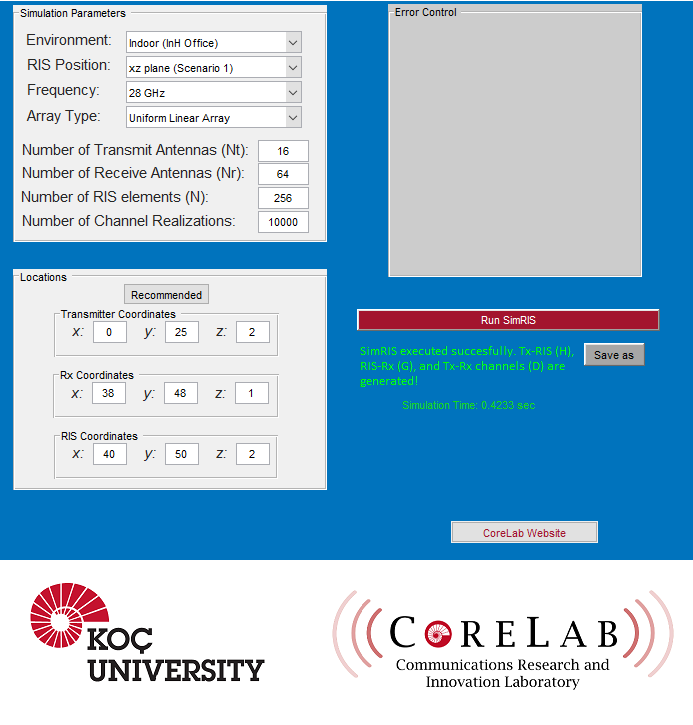}
		\vspace*{-0.1cm}\caption{ Graphical user interface (GUI) of the SimRIS Channel Simulator v2.0.}\vspace*{-0.7cm}
		\label{fig:GUI}
	\end{center}
\end{figure}


In this section, the open-source, user-friendly, and widely applicable \textit{SimRIS Channel Simulator v2.0} is presented. Our SimRIS Channel Simulator opens a new line of research for physical channel modeling of RIS-empowered networks by taking into account the LOS probabilities between terminals, array responses of RISs and Tx/Rx units, RIS element gains, realistic path loss and shadowing models, and environmental characteristics in different propagating environments and operating frequencies. More specifically, it considers Indoor Hotspot (InH) -  Indoor Office and Urban Microcellar (UMi) - Street Canyon environments for popular mmWave operating frequencies of 28 GHz and 73 GHz from the 5G channel model \cite{3GPP_5G}. Compared to its earlier version with only single-input single-output (SISO) Tx/Rx terminals, its v2.0 considers MIMO terminals with  different types of arrays. We support terminals in the far-field of the RIS only, which is a reasonable assumption for mmWaves with shorter wavelengths and smaller RIS sizes, while the integration of near-field RIS models will be evaluated in its future releases.

The considered generic 3D geometry is given in Fig. 2 for the representation of physical channel characteristics. In this setup, the RIS is mounted on the $xz$-plane while a generalization is possible in the simulator. Our model considers various indoor and outdoor wireless propagation environments in terms of physical aspects of mmWave frequencies while numerous practical 5G channel model issues are adopted to our channel model \cite{3GPP_5G}. For a considered operating frequency and environment, the number of clusters ($C$), number of sub-rays (scatterers) per cluster ($S_c$), and the positions of the clusters can be determined by following the detailed steps and procedures in \cite{SimRIS_jour_new}. More specifically, according to the 5G channel model, the number of clusters and scatterers are determined using the Poisson and uniform distributions with certain parameters, respectively. Although the clusters between Tx-RIS and Tx-Rx can be modeled independently, it can be assumed that the Rx and the RIS might share the same clusters when the Rx is located relatively closer to the RIS. 
 Due to the fixed orientation of the Tx and the RIS, the array response vectors of the Tx and the RIS are easily calculated for given azimuth and elevation departure/arrival angles. However, it is worth noting that if azimuth and elevation angles are generated randomly for the Tx, due to fixed orientation of the RIS, they will not be random anymore at the RIS and should be calculated from the 3D geometry using trigonometric identities. Nevertheless, the array response vector of the Rx can be calculated with randomly distributed azimuth and elevation angles of arrival due to the random orientation.

Using the SimRIS Channel Simulator, the wireless channels of RIS-aided communication systems can be generated with tunable operating frequencies, number of RIS elements, number of transmit/receive antennas, Tx/Rx array types, terminal locations, and environments. As discussed earlier, in SimRIS Channel Simulator v2.0, MIMO-aided Tx and Rx terminals are incorporated into its earlier version, and the array response vectors and receiver orientation are reconstructed by considering this MIMO system model. This new version also offers two different types of antenna array configurations: Uniform linear array (ULA) and uniform planar array (UPA), and the corresponding array response vectors are calculated according to the selected antenna array type. The graphical user interface (GUI) of our SimRIS Channel Simulator v2.0 is given in Fig. \ref{fig:GUI}. In this GUI, the selected scenario specifies the RIS position for both indoor and outdoor environments as $xz$-plane (Scenario 1) and $yz$-plane (Scenario 2). Considering the 3D geometry illustrated in Fig. 2, Tx, Rx, and RIS positions can be manually entered into the SimRIS Channel Simulator. Furthermore,  $N_t$ (the number of Tx antennas), $N_r$ (the number of Rx antennas), $N$ (the number of RIS elements), and the number of channel realizations are also user-selectable input parameters and these options offer a flexible and versatile channel modeling opportunity to the users.  Considering these input parameters, our simulator produces $\mathbf{H}$, $\mathbf{G}$, and $\mathbf{D}$ channel matrices in (2) by conducting Monte Carlo simulations for the specified number of realizations under $28$ and $73$ GHz mmWave frequencies. The general expressions of these channel matrices are given in Table 1 for the interested readers. Here, the double summation terms stem from the random number of clusters and scatterers and the LOS components might be equal to zero with a certain probability for increasing distances. 

In a nutshell, our open-source SimRIS Channel Simulator package considers a narrowband channel model for RIS-empowered communication systems for both indoor and outdoor environments and it includes various physical characteristics of the wireless propagation environment.  The open-source nature of our simulator, which is written in the MATLAB programming environment, encourages all researchers to use and contribute to the development of its future versions by reaching a wider audience. We aim a Web-based simulator application and more sophisticated propagation scenarios (wideband and time-varying channels) in its future versions.

\begin{table*}[!t]
\centering
\begin{threeparttable}
\caption{System Configurations and Simulation Parameters For SimRIS Channel Simulator}
\label{tab:parameter}
\begin{tabular}{|c|l|}
\hline
\textbf{}                                           & \begin{tabular}[c]{@{}l@{}}$\mathbf{H}= \gamma \sum\limits_{c=1}^{C} \sum\limits_{s=1}^{S_c} \beta_{c,s} \sqrt{G_e(\theta_{c,s}^{\text{T-RIS}})L_{c,s}^{\text{T-RIS}}} \,\, \mathbf{a} ( \phi_{c,s}^{\text{T-RIS}}, \theta_{c,s}^{\text{T-RIS}} ) \mathbf{a}^T( \phi_{c,s}^{\text{Tx}}, \theta_{c,s}^{\text{Tx}} ) + \mathbf{H}_{\text{LOS}}$\end{tabular} \\ \cline{2-2} 
\textbf{Channel Matrices*:}                           & $\mathbf{G}=
\bar{\gamma} \sum\limits_{c=1}^{\bar{C}} \sum\limits_{s=1}^{\bar{S_c}} \bar{\beta}_{c,s} \sqrt{G_e(\theta_{c,s}^{\text{RIS-R}}) L_{c,s}^{\text{RIS-R}}} \,\, \mathbf{a} ( \phi_{c,s}^{\text{Rx}}, \theta_{c,s}^{\text{Rx}} ) \mathbf{a}^T( \phi_{c,s}^{\text{RIS-R}}, \theta_{c,s}^{\text{RIS-R}} ) + \mathbf{G}_{\text{LOS}}$                                                                                                         \\ \cline{2-2} 
\textbf{}                                           & \begin{tabular}[c]{@{}l@{}}$\mathbf{D}=
\tilde{\gamma} \sum\limits_{c=1}^{\tilde{C}} \sum\limits_{s=1}^{\tilde{S}_c} \tilde{\beta}_{c,s} \sqrt{L_{c,s}^{\text{T-R}}}\mathbf{a} ( \phi_{c,s}^{\text{Rx'}}, \theta_{c,s}^{\text{Rx'}} ) \mathbf{a}^T( \phi_{c,s}^{\text{Tx'}}, \theta_{c,s}^{\text{Tx'}} ) + \mathbf{D}_{\text{LOS}}$\end{tabular}               

\\ \hline
\multirow{2}{*}{\textbf{Environments:}}          & Indoor: InH Indoor Office                                                                                                                                                                                                                                                           \\ \cline{2-2} 
                                                    & Outdoor: UMi Street Canyon                                                                                                                                                                                                                                                          \\ \hline
\multirow{2}{*}{\textbf{Tx, Rx and RIS Positions:}} & Indoor: (0,25,2), (45,45,1), (40,50,2)                                                                                                                                                                                                                                              \\ \cline{2-2} 
                                                    & Outdoor: (0,25,20), (50,50,1), (40,60,10)                                                                                                                                                                                                                                             \\ \hline
\textbf{Operating Frequency:}                       & 28 GHz                                                                                                                                                                                                                                               \\ \hline
\textbf{Antenna Array Type:}                        & \begin{tabular}[c]{@{}l@{}}Uniform linear array (ULA) and uniform planar array (UPA)\end{tabular}                                                                                                                                                                                   \\ \hline

\textbf{Algorithm for Phase Shifts:}                        & \begin{tabular}[c]{@{}l@{}}Pseudoinverse (pinv)-based algorithm \cite{Pinv_2019} \end{tabular}                                                                                                                                                                                   \\ \hline
\end{tabular}
\begin{tablenotes}
      \small
      \item[*]  $\gamma / \bar{\gamma}/ \tilde{\gamma}$: normalization factor \cite{SimRIS_Latincom}, $\beta_{c,s} / \bar{\beta}_{c,s}/\tilde{\beta}_{c,s}$: complex Gaussian distributed gain of the $(c,s)$th propagation path, $G_e(.)$: RIS element gain, $L^i_{c,s}$: attenuation of the $(c,s)$th propagation path \cite{3GPP_5G}, $\mathbf{a} \left( \phi^i_{c,s}, \theta^i_{c,s}\right)$: array response vectors for the considered azimuth ($\phi^i_{c,s}$) and elevation angles ($\theta^i_{c,s}$), $\mathbf{H}_{\text{LOS}}$/ $\mathbf{G}_{\text{LOS}}$/ $\mathbf{D}_{\text{LOS}}$: LOS component of sub-channels ($i$: indicator for the corresponding path/terminal as $i\in \lbrace \text{T-RIS},\text{RIS-R},\text{T-R},\text{Tx},\text{Rx}\rbrace$). 
    \end{tablenotes}
\end{threeparttable}
\end{table*}

\section{Simulation of Indoor/Outdoor Scenarios Using SimRIS Channel Simulator}
\begin{figure}[!t]
	\begin{center}
		\includegraphics[width=1\columnwidth]{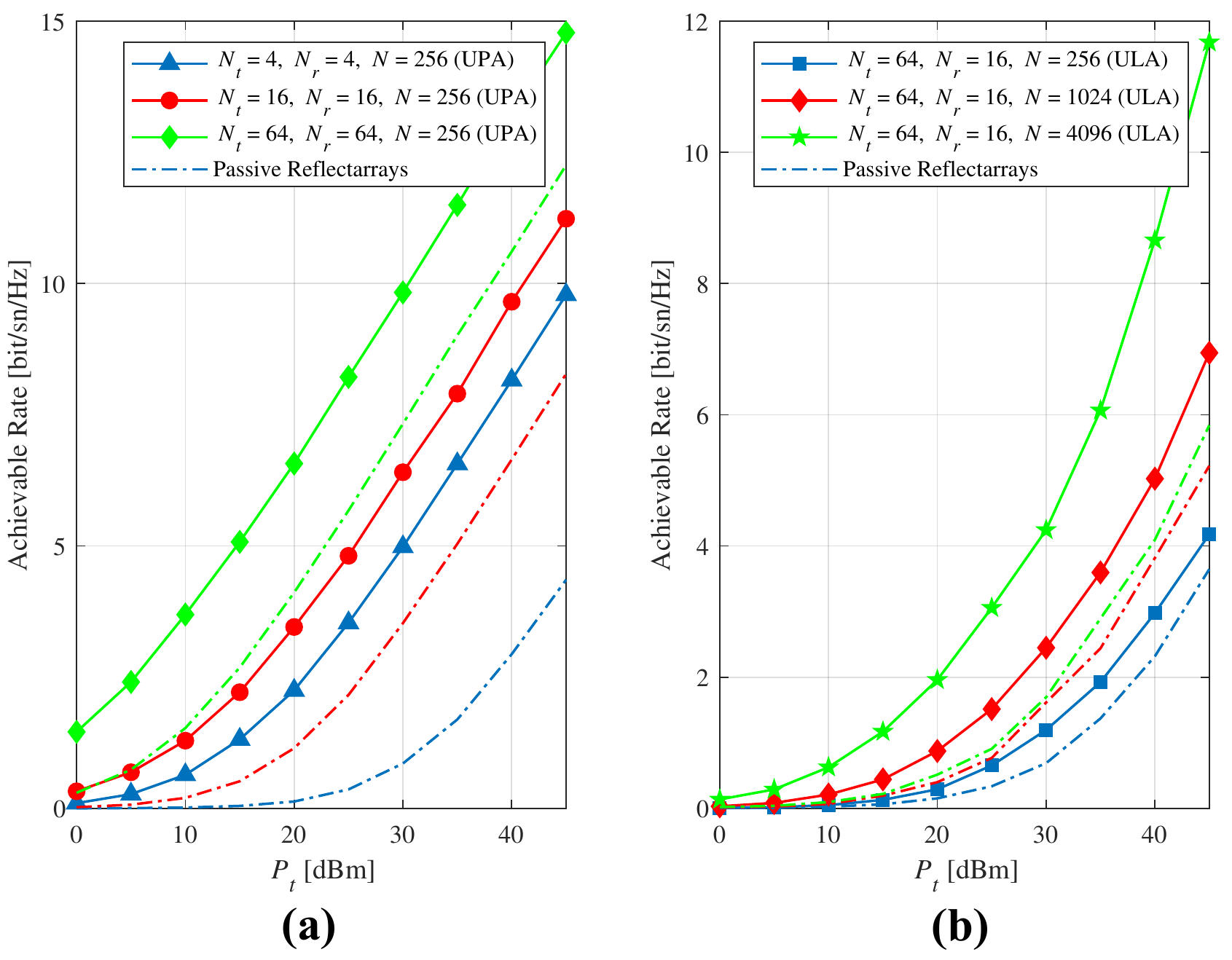}
		\vspace*{-0.3cm}\caption{Achievable rates of RIS-assisted systems (a) in an indoor environment under different $N_t$ and $N_r$ with fixed $N$, and (b) in an outdoor environment under different $N$ with fixed $N_t$ and $N_r$.}\vspace*{-0.5cm}
		\label{fig:Fig3}
	\end{center}
\end{figure}
The goal of the section is to pave the way for the detailed elaboration of how RISs can be effectively used in future networks to enhance the existing transmission systems. In next-generation wireless networks, it is inevitable to use a high number of antennas at Rx and Tx units. Many transmission scenarios involving massive MIMO systems offer significant increases in achievable rates. However, when high frequencies are considered, especially using a large number of antennas in mobile users will increase the cost of signal processing and required hardware. By using RISs in transmission, it will be still possible to obtain high achievable rates, while alleviating the cost of physical implementation and signal processing. The summary of the considered system configurations, generated channel matrices, and computer simulation parameters are given in Table 1, which will be used for the numerical results in this section. Here, all computer simulations are conducted at an operating frequency of $28$ GHz and pseudoinverse (pinv)-based algorithm \cite{Pinv_2019} is used for adapting the phase shifts of an RIS-assisted MIMO transmission system. Here, the selection of the pinv-based algorithm can be explained by its relatively easy to implement architecture without any iterative operations, compared to non-convex optimization algorithms. The ergodic achievable rate for the MIMO systems is given by
\begin{equation}
    R = \log_2\det\left(\mathbf{I}_{N_r}+\dfrac{P_t}{\sigma^2}\mathbf{C}^H\mathbf{C}\right) \, \text{bit/sec/Hz}
\end{equation}
where $\mathbf{I}_{N_r}$ and $\sigma^2$ denote  $N_r \times N_r$ identity matrix and the average noise power, respectively. The noise power is assumed to be $-100$ dBm for all simulations in this section. As stated above, the pinv algorithm can be effectively used to maximize $R$, without resorting to more complicated convex optimization problems due to non-concavity of the MIMO channel capacity \cite{Wu_Tutorial}. Finally, the determination and execution of optimal RIS adaptation algorithms are  beyond the scope of the SimRIS Channel Simulator, at least in this version, while interested readers can easily use our generated channels in their own optimization models.

\begin{figure*}[!t]
	\begin{center}
		\includegraphics[width=1.7\columnwidth]{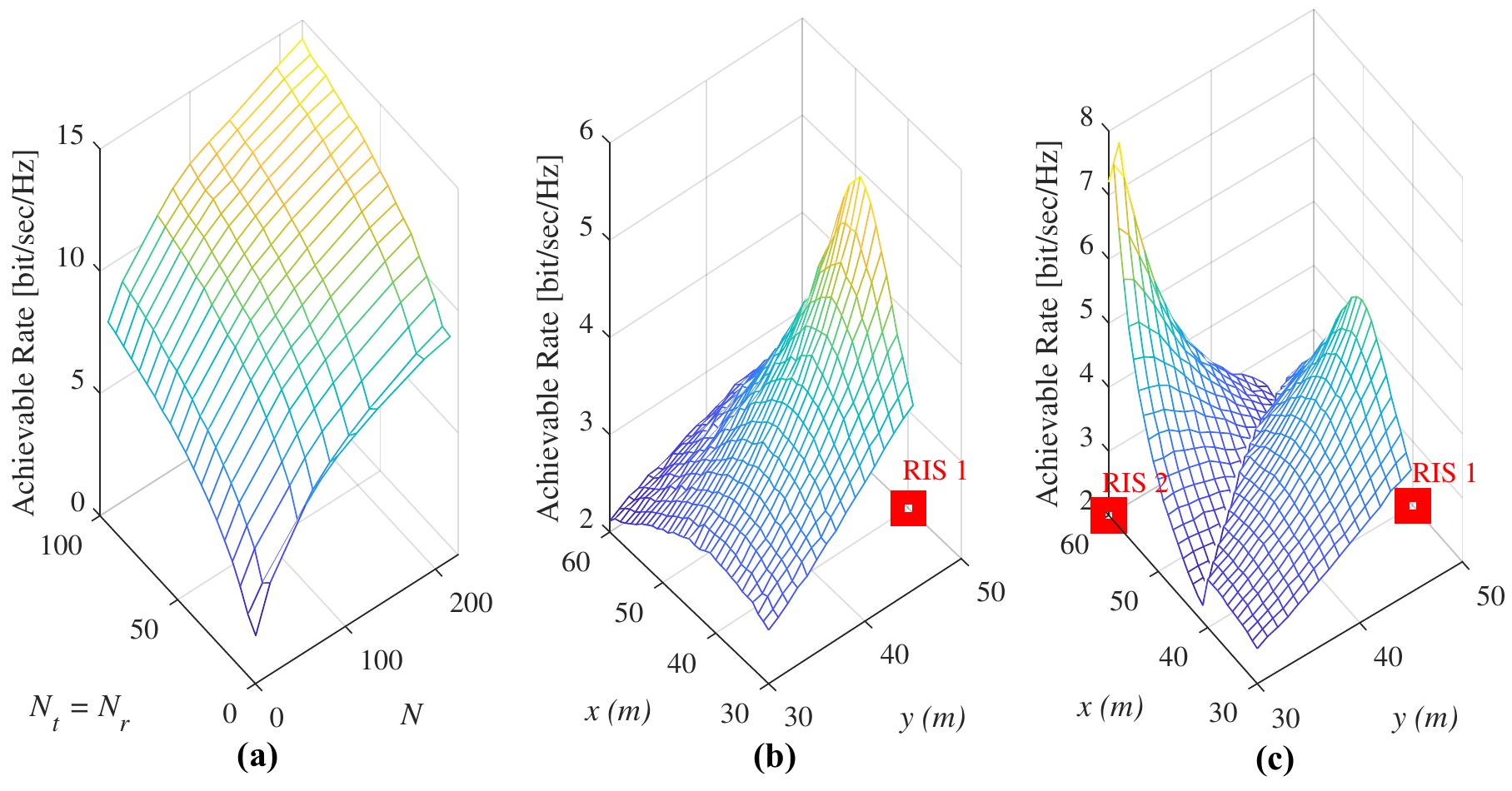}
		\vspace*{-0.3cm}\caption{ Achievable rate analysis using SimRIS Channel Simulator: (a) For varying $N_t$/$N_r$ and $N$, (b) For a single RIS with varying Rx positions, (c) For two RISs with varying Rx positions. Here, the red squares indicate the positions of the RISs on the $x$ and $y$-axes.}\vspace*{-0.5cm}
		\label{fig:3D_result}
	\end{center}
\end{figure*}

In Fig. 4(a) the effect of varying numbers of transmit and receive antennas on the achievable rate is examined, considering the indoor environment parameters in Table 1. Using UPAs both in transmitter and receiver, increasing $N_t$ and $N_r$ values have been observed to enable a significant increase in achievable rates for $N = 64$ under increasing $P_t$. As clearly seen from Fig. 4(a), compared to the passive reflectarrays, which simply reflect the incident waves without any phase adjustment, RISs can provide a remarkable improvement in the achievable rate for the same $N$ value. In Fig. 4(b), the achievable rate is investigated by keeping the number of antennas constant and increasing $N$ considering the outdoor environment parameters in Table 1. As seen from Fig. 4(b), a substantial increase in achievable rate is also obtained by increasing $N$  with fixed $N_t$ and $N_r$ when ULA type antennas are used. Moreover, the RIS-assisted system also reaches a considerably better achievable rate performance than the reference passive reflectarray-based system for the outdoor environment. Here, the RIS is positioned to have a LOS transmission link with the Tx and the Rx, as well as to have a closer RIS-Rx distance for both indoor and outdoor environments when the Tx-Rx path is blocked by obstacles. Figs. 4(a) and 4(b) show that increasing numbers of antennas and reflecting elements have a similarly favorable effect on the achievable rate, while the overall cost of the latter might be much lower due to its passive nature. We also note that the differences in Figs. 4(a) and (b) can be mainly attributed to the longer range transmission in the outdoor environment. Finally, it is worth noting that our channel simulator can be used for the calculation of system throughput by effective signal-to-noise ratio mapping tables.

In order to get a more accurate understanding of the effects of the number of Tx/Rx antennas and reflecting elements, we also observe the joint effect of these two parameters on the achievable rate in Fig. \ref{fig:3D_result}. The joint effect of the number of antennas and reflecting elements on the achievable rate is demonstrated in Fig. \ref{fig:3D_result}(a) for an indoor environment when $N_t = N_r$ and $P_t = 40$ dBm. As seen from Fig. \ref{fig:3D_result}(a), if the number of antennas is doubled under a fixed $N$,  an increase of approximately $2$ bit/sec/Hz is achieved in the achievable rate. Similarly, if $N_t$ and $N_r$ are kept constant, when $N$ is doubled, a gain of about $2$ bit/sec/Hz is also obtained. Since doubling the number of reflectors is much less costly than doubling the number of Tx/Rx antennas, the demand for ultra-reliable and high-speed transmission that will emerge in the next-generation wireless networks can be met effectively by increasing the $N$.

Fig. \ref{fig:3D_result}(b) explores the use of an RIS for indoor coverage extension. When the direct path between the Tx and the Rx is lost due to surrounding objects and obstacles, reliable transmission conditions are guaranteed by placing an RIS near to the Rx. Here, we consider the indoor parameters in Table 1 and the effect of the movement of the Rx on the $xy$-plane to its achievable rate is examined. It is observed that the highest rates are achieved in particular regions, where the distance between the Rx and the RIS is $10$ meters or less and this reveals the promising role of an RIS in future wireless networks while providing a general perspective for the effective positioning of RISs. Under the LOS-dominated Tx-RIS-Rx channels,  the use of an RIS in transmission improves signal quality even in dead zones or cell-edges by providing low-cost and energy-efficient solutions with low complexity. 

If we want to go one step further to provide a wider coverage extension, the idea of using more than one RISs in transmission might be considered in the system design.  The variation of the achievable rate of the Rx for its different locations is observed in Fig. \ref{fig:3D_result}(c) when a second RIS with coordinates $(60,30,2)$  is placed in the system in addition to the first RIS at $(40,50,2)$. Here, we aim to increase the quality of the received signal by adjusting the reflecting element phases of the RIS, which is closer to the Rx. In other words, our two RISs create a handover opportunity for the Rx to boost its signal strength. As seen from Fig. \ref{fig:3D_result}(c), when the Rx is located in areas close to one of the RISs, one can observe a significant increase in its achievable rate. Furthermore, in locations, where the Rx is far from both RISs, rapid decreases are observed in the achievable rate. From the given results, we observe that RISs can be considered as powerful and energy-efficient tools to boost the achievable rate as well as for the coverage extension, and alleviates the large number of antenna requirements in next-generation wireless networks. 

In light of our extensive computer simulations using the SimRIS Channel Simulator, we can certainly state that RISs might have a huge potential under certain conditions such as LOS-dominated Tx-RIS-Rx channels and shorter RIS-Rx distances. 
Another important point here is that an RIS provides the maximum benefit when the direct path is lost due to surrounding objects and obstacles since the presence of a LOS-dominated and a strong Tx-Rx path will guarantee a reliable communication eliminating the need for an RIS. Our findings are also consistent with one of the most recent practical applications of RISs, scatterMIMO \cite{scatterMIMO}, which considers careful RIS size adjustment and positioning to achieve a strong Tx-RIS-Rx composite channel.

%%\begin{figure}[!t]
% 	\begin{center}
% 		\includegraphics[width=0.9\columnwidth]{Fig5.pdf}
% 		\vspace*{-0.3cm}\caption{Achievable rates of RIS-assisted systems for Scenario 1 in indoors under $28$ and $73$ GHz operating frequency.}\vspace*{-0.5cm}
% 		\label{fig:Fig5}
% 	\end{center}
 % \end{figure}

\section{Future Directions}
Despite their huge potential and rich applications for future wireless networks, we believe that there is still a long way to go for the standardization of this promising technology. More importantly, we need to bridge the gap between theoretical analysis and real-world deployments for RIS-empowered systems. For instance, the received signal power given in (1), as well as the achievable rate results obtained by the SimRIS Channel Simulator, can be considered as benchmarks for practical systems due to hardware imperfections, estimation/feedback errors, and the limited phase resolution of practical RIS elements. Furthermore, we do not have clear answers to the total energy consumption and cost of RISs in realistic systems.

Open research issues towards 6G wireless networks are summarized as follows:
\begin{itemize}
\item Practical path-loss/channel modeling and real-time testing of large-scale RISs in different propagating environments 
    \item Determination of convincing use-cases in which the RISs might have a huge potential to boost the communication quality-of-service
    \item Assessment of practical protocols for reconfigurability of RISs
    \item Determination of fundamental performance limits of RIS-assisted networks
    \item Robust optimization and resource allocation issues in space, time, and frequency domains
    \item Optimal positioning of multiple RISs and optimization/coordination of the overall network
    \item Development of EM-based RIS models and exploration of hardware imperfections/effects
    \item Exploration of effective mmWave and Terahertz communication systems with RISs
    \item Exploration of the potential of RISs for beyond communication (sensing, radar, localization etc.)
    \item AI-driven tools for designing/optimizing/reconfiguring surfaces
    \item Exploration of futuristic scenarios such as very high number of devices, combination of enhanced mobile broadband and ultra-reliable and low latency communications, and very high mobility
    \item Standardization and integration into existing wireless communication networks (5G, 6G, IoT, IEEE 802.11x), which requires a joint effort from academia and industry.
    
\end{itemize}

\section{Conclusion}
It is inevitable that 6G networking will necessitate radically new concepts for the PHY design of next-generation radios. Within this context, this article has discussed the potential of intelligent surfaces-empowered communication by following a bottom-up approach from its main methodology and use-cases to its open problems and potential future directions. Special emphasis has been given to the open-source and widely acceptable SimRIS Channel Simulator, which considers a novel physical channel modeling methodology for RIS-empowered systems operating in different environments and frequencies. Based on our initial findings and numerical results using this simulator, from a communication theoretical perspective, we can certainly state that RISs might be game changers after careful assessment and planning for certain use-cases and environments. We have also discussed the future challenges to bridge the gap between theory and practice for RIS-empowered systems. The future of wireless communications looks very exciting and stay tuned into this promising technology!

\section*{Acknowledgment}

The work of E. Basar was supported in part by the Scientific and Technological Research Council of Turkey (TUBITAK) under Grant 120E401. 
\ifCLASSOPTIONcaptionsoff
  \newpage
\fi

\begin{IEEEbiography}
 [{\includegraphics[width=1in,height=1.25in,clip,keepaspectratio]{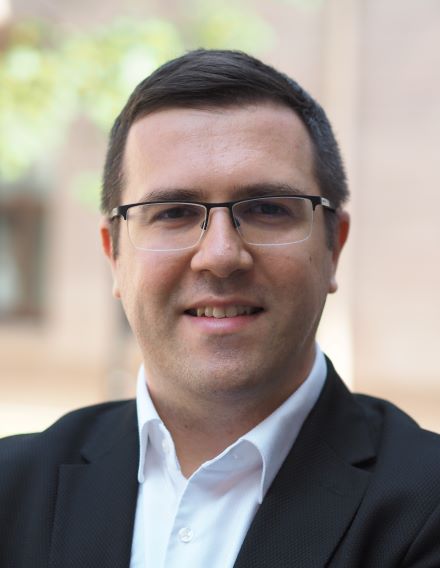}}]{Ertugrul Basar} (Senior Member, IEEE) received the B.S. degree (Hons.) from Istanbul University, Turkey, in 2007, and the M.S. and Ph.D. degrees from Istanbul Technical University, Turkey, in 2009 and 2013, respectively. He is currently an Associate Professor with the Department of Electrical and Electronics Engineering, Ko\c{c} University, Istanbul, Turkey and the director of Communications Research and Innovation Laboratory (CoreLab). His primary research interests include MIMO systems, index modulation, intelligent surfaces, waveform design, visible light communications, and signal processing for communications. Dr. Basar currently serves as a Senior Editor of the \textsc{IEEE Communications Letters} and the Editor of the \textsc{IEEE Transactions on Communications}, and \textit{Frontiers in Communications and Networks}.
\end{IEEEbiography}

\begin{IEEEbiography}
[{\includegraphics[width=1in,height=1.25in,clip,keepaspectratio]{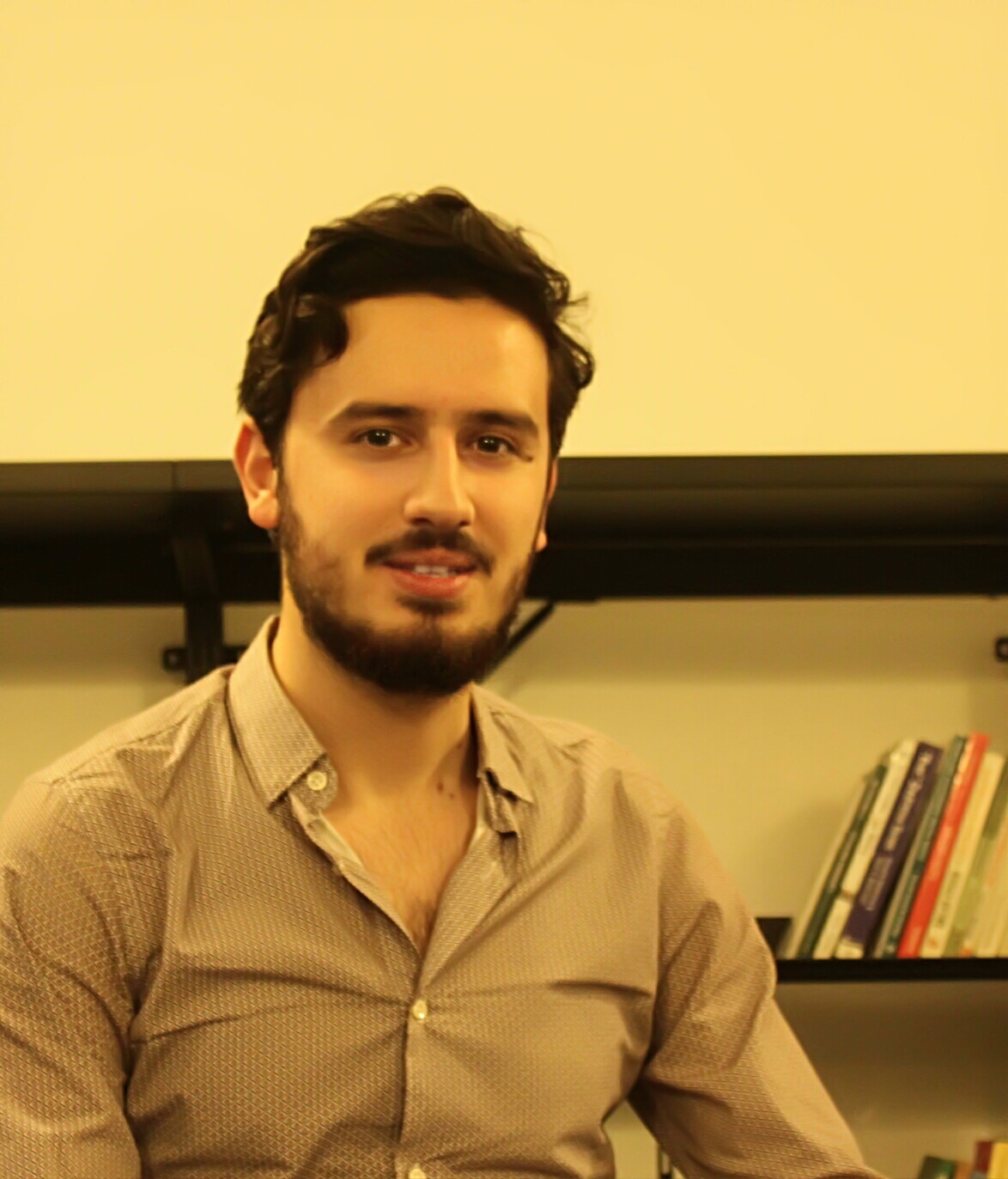}}]{Ibrahim Yildirim} (Graduate Student Member) received the B.S. and M.S. degrees (Hons.) from Istanbul Technical University, Turkey, in 2017 and 2019, respectively. He is currently pursuing the Ph.D. degree at Ko\c{c} University. He is also a Research and Teaching Assistant at Istanbul Technical University. His current research interests include MIMO systems, index modulation, and intelligent surfaces. He has been serving as a Reviewer for the \textsc{IEEE Journal on Selected Areas in Communications}, the \textsc{IEEE Transactions on Vehicular Technology}, and the \textsc{IEEE Communications Letters}.

\end{IEEEbiography}

%\newpage


%\vfill


\bibliographystyle{IEEEtran}
\bibliography{bib_2020}

\end{document}